\documentclass[conference]{IEEEtran}
\IEEEoverridecommandlockouts
\usepackage{cite}
\usepackage{amsmath,amssymb,amsfonts}
\usepackage{algorithmic}
\usepackage{graphicx}
\usepackage{textcomp}
\usepackage{xcolor}
\usepackage{booktabs}
\usepackage{flushend}
\usepackage{soul}
\usepackage{balance}
\usepackage{hyperref}

\newcommand{\R}{\mathbb{R}}
\usepackage{subcaption}
\usepackage{multirow}
\begin{document}

\title{
A Comparative Study  of Transformer-based Neural Text Representation Techniques on Bug Triaging
}

\author{\IEEEauthorblockN{Atish Kumar Dipongkor}
\IEEEauthorblockA{\textit{Dept. of Computer Science} \\
\textit{University of Central Florida}\\
Orlando, USA \\
akd@knights.ucf.edu}
\and 

\IEEEauthorblockN{Kevin Moran}
\IEEEauthorblockA{\textit{Dept. of Computer Science} \\
\textit{University of Central Florida}\\
Orlando, USA \\
kpmoran@ucf.edu}
}

\maketitle

\thispagestyle{plain}
\pagestyle{plain}
\begin{abstract}
Bug report management has been shown to be an important and time consuming software maintenance task. Often, the first step in managing bug reports is related to triaging a bug to the appropriate developer who is best suited to understand, localize, and fix the target bug. Additionally, assigning a given bug to a particular part of a software project can help to expedite the fixing process. However, despite the importance of these activities, they are quite challenging, where days can be spent on the manual triaging process. Past studies have attempted to leverage the limited textual data of bug reports to train text classification models that automate this process -- to varying degrees of success. However, the textual representations and machine learning models used in prior work are limited by their expressiveness, often failing to capture nuanced textual patterns that might otherwise aid in the triaging process. Recently, large, transformer-based, pre-trained neural text representation techniques (i.e., large language models or LLMs) such as BERT and CodeBERT have achieved greater performance with simplified training procedures in several natural language processing tasks, including text classification. However, the potential for using these techniques to improve upon prior approaches for automated bug triaging is not well studied or understood.

Therefore, in this paper we offer one of the first investigations that fine-tunes transformer-based language models for the task of bug triaging on four open source datasets, spanning a collective 53 years of development history with over 400 developers and over 150 software project components. Our study includes both a quantitative and qualitative analysis of effectiveness. Our findings illustrate that DeBERTa is the most effective technique across the triaging tasks of developer and component assignment, and the measured performance delta is statistically significant compared to other techniques. However, through our qualitative analysis, we also observe that each technique possesses unique abilities best suited to certain types of bug reports.
\end{abstract}


\section{Introduction} 
Complex software systems have several modules or components and can experience a large number of bugs during their life-cycle. For example, according to Avnik \textit{et al.}~\cite{anvik2006should}, nearly 300 bugs are found in different Mozilla projects every day. In the Firefox project alone, an average of eight bugs are reported daily that need to be triaged. Similarly, the Eclipse project receives 37 bug reports every day~\cite{anvik2006automating}. To resolve these bugs, the first maintenance task required for addressing an incoming report is \textit{triaging} the bug, or associating it with the general module or \textit{component} affected and/or the \textit{developer} who is best equipped to localize and fix the underlying fault. Bug triaging for a large-scale software system is not straightforward, is prone to human errors and inconsistencies, and often delays bug resolution~\cite{crowston2006information,jeong2009improving}.

In complex open-source software products, a large number of developers are involved and each developer possesses a different set of skills and knowledge. Given the complexity of modern software and sheer amount of expertise required to effectively assign bugs to proper developers/components, the task is generally considered to be cognitively challenging and prone to errors~\cite{zhang2016literature}. For instance, 233 developers are involved in the Mozilla Core project which makes it challenging to identify whether a particular developer has the appropriate knowledge to fix a given bug. Likewise, assigning components or modules to bug reports is also not a trivial task. Large software products have numerous components, with Mozilla Core alone consisting of 105 components. As a result, a human triager must be familiar with each component if (s)he wants to properly assign them to a reported bug. The challenging nature of triaging means that this is often performed manually, and generally consumes a considerable amount of time. For instance, in Eclipse and Mozilla, it takes about 40 and 180 days, respectively, to assign a bug to a developer~\cite{bhattacharya2012automated}. Moreover, in some projects like ArgoUML and PostgreSQL, the median time-to-fix for bugs can be around 200 days~\cite{kim2006long}. 

To address the challenges of manual bug triaging, researchers have proposed automated Machine Learning (ML)~\cite{xuan2017automatic,yang2014towards} and Deep Learning (DL)~\cite{lee2017applying,mani2019deeptriage} methods that can learn from past data to recognize and triage incoming bugs. These methods treat bug triaging as a multi-class text classification problem, where the bug's title and description serve as textual data, and the specific developer and component are the labels or classes. One of the major limitations of existing ML and DL techniques is related to the semantic representation of text, as there is often nuanced patterns captured in the natural language and code contained within bug reports that make it challenging to properly classify such artifacts among large and diverse sets of developers and components. For example, past ML-based approaches utilize word-frequency based text representations like TF-IDF (term-frequency - inverse document frequency), which represent text by measuring the frequency of isolated terms. However, the \textit{ordering} of words may reveal certain patterns in text that are not captured by such term representations. Word embedding techniques such as Word2Vec~\cite{church2017word2vec} and ELMo~\cite{PetersNIGCLZ18} have been adopted to address this issue by incorporating word order or context awareness~\cite{zaidi2020applying,lee2017applying,mani2019deeptriage}. However, even these word embedding techniques are limited in that they generate a \textit{fixed embedding} for each word which cannot be updated during training for downstream tasks~\cite{PetersNIGCLZ18}, such as bug triaging, or other Natural Language Understanding (NLU) tasks. Updating the weights of the pre-trained model during fine-tuning allows the model to adapt to the specific patterns and features of the task at hand. For example, if the downstream task is sentiment analysis, the fine-tuning process can adjust the weights of the pre-trained model to better capture the sentiment-related features in the task-specific data~\cite{devlin2018bert}.

Recently, transformer-based neural text embedding techniques such as BERT~\cite{devlin2018bert} and its variations have shown remarkable performance in NLU tasks. The Transformer architecture's self-attention mechanism can capture semantic information about the meaning of a document at a higher level than previous NLP techniques as it allows for the model to focus on relevant parts of the input sequence and capture long-range dependencies between words. Additionally, the pre-trained weights of these models can be fine-tuned for any down-stream task, often resulting in better performance on such tasks as compared to training a model from scratch. This is due to the fact that pre-trained models gain a rich understanding of general language semantics and syntax from self-supervised training procedures (e.g., next token prediction or masked language modeling (MLM)) and only need to be adapted to the specifics of the task. However, the software engineering research community currently does not have a complete understanding of how these techniques perform when applied to the task of bug triaging. Thus, in this paper, we provide findings from a comprehensive empirical study that applies six recent transformer-based language models to bug triaging, shedding light on this understudied aspect of bug triaging and offering promising directions for future work at the intersection of large language models (LLMs) and bug report management.

Our study context consists of 136k bug reports collected from a combined 53 years of maintenance history from four popular open source software repositories. These repositories consist of 165 distinct software components to which a combined 462 developers have made contributions. We study 6 neural transformer-based language models (each a variation of the popular BERT architecture~\cite{devlin2018bert} with different key properties) and one frequency-based (TF-IDF + SVM) based baseline approach. More specifically we examine ALBERT, BERT, CodeBERT, DeBERTa, DistilBERT and RoBERTa. To carry out our comparative study we adopt the sequence classification strategy of BERT to fine-tune all of our studied neural-based models to assign developers and components to bug reports. Then, we conduct both quantitative and qualitative analyses of various effectiveness metrics, and use statistical tests to investigate whether one model outperforms another to a statistically significant degree. We also conduct an orthogonality analysis to better understand the unique behavior of each technique, and an error analysis to investigate the underlying reasons for common failures. Our results capture several significant findings that further our understanding of how transformer-based LMs perform when applied to the task of bug triaging, and point toward important directions for future investigation. We make all of our code and data available in an online appendix~\cite{appendix} to facilitate the replication and reproducibility of our work, and to encourage future research on adapting neural transformer-based LMs to the task of bug triaging. Below we summarize our findings:

\begin{itemize}
    \item DeBERTa performs significantly better than other transformer-based language models in both developer and component assignment tasks;
    \item{Somewhat surprisingly, the simpler TF-IDF-based SVM baseline performs best, for the task for developer assignment, on two of our four studied datasets, illustrating that a well-tuned term frequency-based approach can perform well when triaging bugs to developers;}
    \item Each technique has a certain degree of orthogonality which indicates that the unique properties of each technique allow them to capture a different set of patterns from bug reports;
    \item Similarity between components and developers hampers the performance of all transformer-based language models. There is a positive correlation between these similarities and number of misclassifications.
\end{itemize}

The rest of this paper is organized as follows. Section \ref{related-work} describes existing works which are closely related to this study. Our methodology and research questions (RQ) are presented in Section \ref{methodology}. In section \ref{experiment}, we present our findings and answer the RQs. Finally, our study is concluded in the Section~\ref{conclusion} with future directions.

\section{Related Work}\label{related-work}
The field of automatic bug triaging is constantly evolving and has seen the use of both ML and DL techniques to triage bug reports. Both types consider bug triaging as a supervised classification problem, where the output classes are the names of the developers, and the training data consists of text from bug reports. However, the primary contrast between them is that ML techniques are trained using hand-crafted features while DL techniques learn their features during training. In the upcoming sub-sections, we categorize the existing studies and discuss it in detail.

\subsection{Automatic Bug Triaging using ML Techniques}
Na\"ive Bayes, Bayesian Networks, C4.5, SVM, Decision Tree (DT), $k$-Nearest Neighbors ($k$NN) \cite{xia2013accurate} and Logistic Regression are the most commonly used algorithms for bug triaging. To train these techniques, various type of features are used. For instance, some used TF-IFD representation of the bug reports only while other used categorical features along with the TF-IFD representation. In addition, some works applied feature selection and extraction techniques on the TF-IFD representation of the bug report using Latent Semantic Indexing (LSI), Latent Dirichlet Allocation (LDA), Principal Component Analysis (PCA) \cite{nath2021principal,nguyen2014topic}. The motivation of using feature selection algorithms is to train models using useful features.

Cubranic and Murphy \cite{murphy2004automatic} are credited with proposing ML algorithms for bug triaging first. They utilized bag of words to represent bug reports and trained a Na\"ive Bayes classifier for the triaging task. However, due to the low accuracy (30\%) of Na\"ive Bayes, Avnik et al. \cite{anvik2006should, anvik2011reducing} conducted a subsequent study. They filtered out noisy data based on bug status and introduced two additional algorithms, namely SVM and DT C4.5. According to their findings, SVM outperformed the other two algorithms on three open source datasets such as Eclipse, Firefox and GCC.

Sarkar \textit{et al.} \cite{sarkar2019improving} conducted an empirical study on a large-scale industry project comparing SVM, Na\"ive Bayes, and Logistic Regression. They trained these models using TF-IDF vectorization of textual descriptions and categorical features of bug reports, and found that Logistic Regression achieved the highest accuracy. In contrast, Lin \textit{et al.} \cite{lin2009empirical} reported that SVM can predict bug assignments and achieve accuracy close to human triager in a different proprietary project.

Ahsan \textit{et al.} \cite{ahsan2009automatic} explored the use of LSI for reducing the dimension of TF-IDF vectors, and achieved the best results using SVM. Nasim \textit{et al.} \cite{nasim2011automated} used Alphabet Frequency Matrix (AFM) with various ML algorithms and found that SVM or its variants perform the best in automatic bug triaging. Similarly, Florea \textit{et al.} \cite{florea2017spark} used LDA for dimensionality reduction, and concluded that SVM outperforms other ML algorithms.

After examining prior research more closely, it is evident that SVM outperforms other ML algorithms on open source projects. In addition to this motivation, the work of Fu \textit{et al.} \cite{fu2017easy} drives us to choose SVM as one of our baselines. They reported that in certain cases, simple ML methods like SVM can learn more efficiently and deliver superior results compared to DL algorithms for automating software engineering tasks. To this end, we trained and compared SVM with transformer-based neural text embedding techniques.

\subsection{Automatic Bug Triaging using DL Techniques}
Using DL for bug triaging, Convolutional Neural Network (CNN) and Recurrent Neural Network (RNN) based techniques are the common ones which use Long Short-Term Memory (LSTM) based text embedding instead of manual feature engineering. Lee \textit{et al.} \cite{lee2017applying} applied Word2Vec \cite{Mikolov2013} word embedding to train a CNN based classifier for bug triaging. Similar to the previous one, Guo \textit{et al.} \cite{guo2020developer} applied developer-activity based CNN techniques (CNN-DA) where they also used Word2Vec for word embedding and applied word segmentation, stop word removal and stemming techniques in their pre-processing step. Mani \textit{et al.} \cite{mani2019deeptriage} used deep bidirectional recurrent neural network with attention (DBRNN-A) to handle long word sequences. Zaidi \textit{et al.}\cite{zaidi2020applying} compared three word embedding such as Word2Vec\cite{Mikolov2013}, GloVe\cite{PenningtonSM14} and ELMo\cite{PetersNIGCLZ18} with CNN to triage the bugs of four different open source projects. They concluded that context-sensitive word embedding, ELMo outperforms the other two. 

Unlike LSTM-based text embedding techniques, transformer-based text embedding techniques are rarely studied for bug triaging. Lee \textit{et al.}~\cite{lee2022light} proposed Light Bug Triage framework by compressing RoBERTa using knowledge distillation \cite{bucilu2006model}. Although they claimed that their framework can prevent the catastrophic forgetting \cite{mccloskey1989catastrophic} issue of large language models (LLM), they did not compare it with any LLM like \cite{devlin2018bert}, \cite{he2020deberta} or \cite{liu2019roberta}.  

\begin{figure*}[tb]
    \centering
    \includegraphics[width=\linewidth]{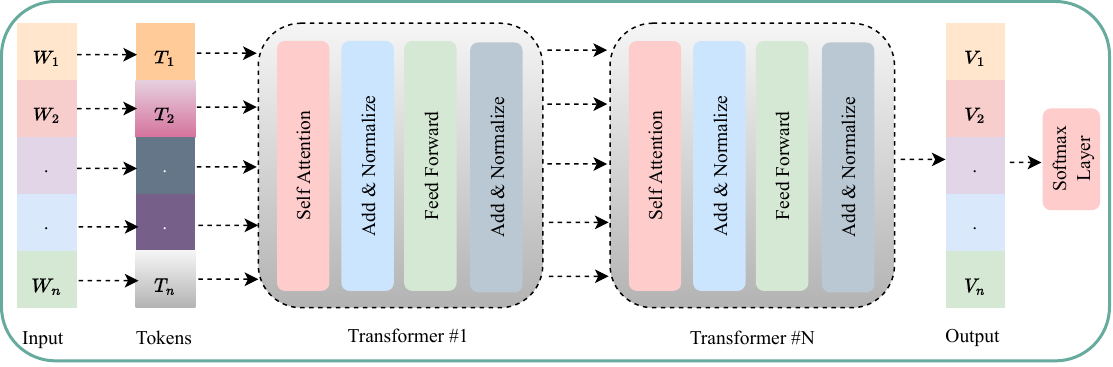}
    \caption{Architecture for fine-tuning Transformer-based Models}
    \label{fine-tuning-arch}
\end{figure*}

\section{Methodology}\label{methodology}
In order to understand the efficiency and effectiveness of our baselines, we formulate the following research questions. The upcoming subsections outline the methodologies we use to address these questions.
\begin{itemize}
\item \textbf{RQ$_1$}: \textit{How well do Transformer-based Neural Text Embedding Techniques perform for automated bug report assignment to developers?}
\item \textbf{RQ$_2$}: \textit{How well do Transformer-based Neural Text Embedding Techniques perform for automated bug report assignment to project components?}
\item \textbf{RQ$_3$}: \textit{What degree of orthogonality exists between our studied techniques?}
\item \textbf{RQ$_4$}: \textit{Why do we observe the effectiveness of the best performing techniques?}
\end{itemize}

\subsection{Dataset Description}

In our study, we make use of the bug report datasets originally collected by Hongrun \textit{et al.} \cite{wu2018empirical}, which contain 200,000 and 220,000 fixed bug reports from the Eclipse and the Mozilla projects, respectively. These projects contain bug reports from the several products of Eclipse and Mozilla; among those we consider four products including: Eclipse JDT, Eclipse Platform, Mozilla Core and Mozilla SeaMonkey. We choose these products because they contain a higher number of bug reports compared to others. A bug report of these datasets consist of crucial details such as the bug's `Summary', `Description', `Assignee', `Component', `Status' and `Timestamp'. To train the benchmark models, we use the Summary' and Description' as input, whereas the Assignee' and `Component' serve as output when necessary. While setting the ground truth, we consider bugs with status as `VERIFIED' and `FIXED'. In addition, we only consider active developers in our experiments. Overall, we follow the same data extraction strategy as other prior works \cite{zhang2020efficient,xuan2012developer,mani2019deeptriage}. Table \ref{dataset-summary} provides a summary of our datasets which are used for our training and evaluation.

\subsection{Preprocessing and Train-Test Split}
To preprocess our training and test data, we adopt similar strategies as those used in previous works of this domain \cite{zhang2020efficient,sarkar2019improving}. Our preprocessing steps include tokenization, removal of stop words, and stemming. To improve the quality of our dataset further, we also eliminate several sources of noise, including hyperlinks, newlines, and special characters. Additionally, we exclude bug reports with fewer than 10 words in their `Summary' + `Description'.

When performing train-test splits, we treat our datasets as time-series data, with the aim of training our models using past data and evaluating using future data. To achieve this, we first sort the bug reports for each product in chronological order based on their `Timestamp'. Next, we divide each dataset into eleven equal folds. During the first run, we train a particular model using fold 1 and test it using fold 2. In the second run, we use both fold 1 and 2 for training and fold 3 for evaluation. We repeat this process up to ten runs, where the overall idea is to train models using $n$ fold(s) and test on the $(n+1)$-th fold. This approach has also been used in previous work \cite{sarkar2019improving,bhattacharya2010fine}. Finally, we evaluate a model based on the average of its ten runs performance.

\subsection{Studied Transformer Text Representation Techniques}

Since the introduction of BERT in 2018, subsequent models have been developed with various improvements to the learned textual representation that are oriented toward different NLU tasks. A core question of our study relates to the differing impact of varying textual representations learned by different types of BERT-like models. DistilBERT~\cite{sanh2019distilbert} reduces the number of parameters a BERT model by 40\% while retaining a reported 97\% of its language understanding capabilities and exhibiting 60\% faster inference and fine-tuning performance. Exploring this model allows us to examine how optimized models perform in our bug triaging task. RoBERTa~\cite{liu2019roberta} improves the performance of BERT by applying dynamic masking to mask language modeling training objective, removing next sentence prediction, and was trained on a larger dataset of text. ALBERT~\cite{lan2019albert} reduced the size of BERT by factorizing the embedding parameters and sharing all parameters across layers. DeBERTa~\cite{he2020deberta} introduced the disentangled self-attention and a new decoding strategy to improve the performance of BERT, and it has been shown to handle long sequences of text more efficiently than others. CodeBERT~\cite{feng2020codebert} was pre-trained to understand both natural language (NL) and programming language (PL), while others were pre-trained only for the former. In this study, we fine-tune all of these techniques for bug triaging and conduct a comparative analysis.

\begin{table}[tb]
\scriptsize
\centering
\setlength{\tabcolsep}{3.7pt}
\caption{Summary of the Dataset}
\label{tbl:dataset}
\label{dataset-summary}
\begin{tabular}{l|l|l|l|l}
\hline
\textbf{Dataset}          & \textbf{Collection Period} & \textbf{Components} & \textbf{Developers} & \textbf{Bug Reports} \\ \hline
Eclipse JDT      & 12 Years          & 6          & 41         & 20565       \\ 
Eclipse Platform & 12 Years          & 21         & 108        & 37240       \\ 
Mozilla Core     & 16 Years          & 105        & 233        & 64238       \\ 
Mozilla SeaMonkey & 13 Years          & 33         & 80         & 14312       \\ \hline
\end{tabular}
\end{table}

\subsection{Benchmark Model Fine-Tuning and Training}

\begin{table*}[tb]
\centering
\setlength{\tabcolsep}{3pt}
\caption{Summary of the Transformer-based models}
\label{bert-summary}
\begin{tabular}{l|l|p{5cm}|l|l}
\hline
\textbf{Model}    & \textbf{Purpose}                                           & \textbf{Pre-training Corpus}   & \textbf{\# Layers} & \textbf{\# Parameters} \\ \hline
BERT (2018)       & Pre-training for NLP tasks                                 & BooksCorpus, English Wikipedia & 12            & 110M                  \\ 
DistilBERT (2019) & Smaller and faster version of BERT                         & Same as BERT & 6                 & 66M                   \\ 
RoBERTa (2019)    & Improved pre-training and data processing compared to BERT & Same as BERT, CC-News, OpenWebText & 12        & 125M                  \\ 
ALBERT (2019)     & Model compression without sacrificing performance          & Same as BERT & 12        & 11M           \\ 
DeBERTa (2020)    & Improved handling of long input sequences                  & Same as BERT & 12        & 163M         \\ 
CodeBERT (2020)   & Pre-trained on both programming and natural language       & GitHub Repositories            & 12             & 209M                  \\ \hline
\end{tabular}
\end{table*}

In this study, we fine-tune the pre-trained transformer-based models for developer and component assignment, while we train the SVM from scratch for the same tasks.

To fine-tune all transformer-based models, we adhere to the standard sequence classification approach, which involves appending a classification layer on top of transformers' outputs. Given the similarity in the overall architecture of the transformer-based models, we utilize it for fine-tuning. The overall architecture works as follows. Initially, tokenization of the input sequence is done using a tokenizer. Then, a special classification token [CLS] is inserted at the sequence's beginning, followed by all tokens passed through multiple transformer layers. Here, it is worth noting that the number of transformer layers employed in each model distinguishes them from one another, and the hyperparameters' settings vary accordingly. A summary of our transformer-based models is provided in Table \ref{bert-summary}. The last transformer layer in these models generates contextualized embeddings for each token, capturing the token's context within the entire sequence. For developer and component classification, only the [CLS] embeddings are considered since it is used for other sequence classification \cite{devlin2018bert}. The [CLS] embedding is commonly known as the aggregated representation of the entire input sequence and is therefore used for sequence classification. Once the [CLS] embedding is obtained from the last layer, it is fed into a softmax classification layer for assigning developers and components to bug reports. Figure \ref{fine-tuning-arch} provides a schematic of the overall fine-tuning architecture for all transformer-based models. We use cross-entropy to minimize loss during fine-tuning since it performs better for unbalanced data, and the AdamW optimizer to update our dense layer's weights. We fine-tune each model for fifteen epochs using a batch size of 32. The PyTorch \cite{pythorch} and Transformer\cite{transformer} libraries are used to implement the fine-tuning architecture, and our source code is accessible via our online appendix~\cite{appendix}.

In contrast to previous works on bug triaging \cite{anvik2006should, anvik2011reducing,nasim2011automated}, we employ a distinct approach in training an SVM classifier. Our method begins with tokenizing bug reports and generating $n$-grams, limiting the selection to $1$-gram and $2$-gram as previously recommended \cite{anvik2006should, anvik2011reducing,nasim2011automated}. We note that certain $n$-grams are frequently found in some bug reports while others are rarely seen. To address this, we apply TF-IDF vectorization to assign significance to the most significant $n$-grams. This results in the creation of a document-term matrix, $M_{br} \in \R^{r \times t}$, where the rows correspond to the number of training documents and the columns to the number of unique terms or $n$-grams. For each $r$-th bug report containing the $t$-th $n$-gram, $M_{br}[r,t]=S_{tf-idf}$, where $S_{tf-idf}$ is the TF-IDF score of the $t$-th $n$-gram. Prior works generally trained SVM classifiers on this TF-IDF representation of bug reports. In contrast, we conduct additional processing of the TF-IDF representation by utilizing Singular Value Decomposition (SVD) to reduce the dimensionality of the matrix. SVD applied to $M_{br} \in \R^{r \times t}$ generates a lower-dimensional matrix, $M_{br}^\prime \in \R^{r \times t^\prime}$, where $t^\prime$ is smaller than $t$ and explains the majority of the variance in the original matrix $M_{br} \in \R^{r \times t}$. Notably, the $t^\prime$ elements are orthogonal to each other, ensuring their independence and lack of redundancy. In our study, we select $t^\prime$ to explain 95\% variance of the original matrix $M_{br}$. Once $M_{br}^\prime \in \R^{r \times t^\prime}$ is derived from a particular training dataset, we train an SVM with a linear kernel to predict developer and components. To implement SVM, we use the scikit-learn \cite{sklearn} python package. Our source code can be found in our online appendix\cite{appendix}.

\subsection{\textbf{RQ}$_1$ \& \textbf{RQ}$_2$: Performance Analysis}
In order to evaluate developer prediction, we utilize two metrics: Top@K Accuracy and Mean Reciprocal Rank (MRR). Top@K Accuracy refers to the percentage of test cases for which the ground truth label appears in the predicted top K candidate list. MRR, on the other hand, is used to evaluate the performance of a ranked list of items, and is defined as the average of the reciprocal ranks of the first relevant instance in each ranked list. Both of these metrics are widely used in recommender systems, and applying them to automated bug triaging systems provides a more realistic view of model effectiveness. For instance, let's assume that a specific model predicts developer $X$ for solving a bug. However, developer $X$ might already be occupied with other projects, which could delay the bug triaging process. In such cases, ranking and suggesting additional developers could expedite the triaging process. To evaluate component prediction, we utilize three metrics: Precision, Recall, and F1-score. Precision measures the proportion of correct predictions made by the model out of all the positive predictions it made. Recall measures the proportion of correct predictions made by the model out of all the actual positive instances. F1-score is the harmonic mean of precision and recall and provides a single score that balances the two metrics. In general, a higher precision score means fewer false positives, a higher recall score means fewer false negatives, and a higher F1-score means better overall performance. In our datasets, we found that each bug report is assigned to a single component only. Therefore, suggesting multiple components or evaluating models using Top@K Accuracy and/or MRR does not make any sense.

We compare different models by taking the average of MRR and F1-score obtained from $n$-fold experiments. Furthermore, we conduct hypothesis testing using paired t-test \cite{lilja2005measuring} to determine whether there is a statistically significant performance difference between the models. For comparing any two of our baselines, our null hypothesis ($H_0$) is there is no significant difference between the average performance ($\mu_0 = \mu_1$) of the two baselines. On the other hand, our alternative hypothesis is, there is a significant difference between the average performance ($\mu_0 \ne \mu_1$) of the two baselines. Then, we calculate the difference between the two models for each pair of folds as follows. Let us assume that ${a_1, a_2, \dots, a_{n}}$ represents the scores of model $A$ from $n$-fold experiments, and ${b_1, b_2, \dots, b_{n}}$ represents the scores of model $B$. We can then calculate the fold-wise differences as ${d_1 = a_1 - b_1, d_2 = a_2 - b_2, \dots, d_{n} = a_{n} - b_{n}}$. After that, we calculate the t-statistic (Eqn. \ref{t-stat}) for the mean performance difference as given below.

\begin{equation}
   \bar{d} = \frac{d_1 + d_2 + \dots + d_{n}}{n}
\end{equation}

\begin{equation}
    \sigma = \sqrt{\frac{1}{n}\sum_{i=1}^{n}(d_i - \bar{d})^2}
\end{equation}

\begin{equation}\label{t-stat}
    t-statistic = \frac{\bar{d}}{\frac{\sigma}{\sqrt{n}}}
\end{equation}

Finally, we determine the p-value with $n-1$ degree of freedom for our measured t-statistic using t-distribution \cite{mendenhall2013introduction}. If the p-value is less than level of significance ($\alpha$), then we reject the null hypothesis. Otherwise, we conclude that there is a significant difference between two models.

\subsection{\textbf{RQ}$_3$: Degree of Orthogonality}
The degree of orthogonality of the baselines refers to how independent these are from each other in terms of assigning developer or component. In other words, the degree of orthogonality measures how much these share in common with each other. For example, MRR and F1-score of two baselines can be equivalent but they may perform better for a different set of bug reports or both may perform better for the same set of bug reports. To that end, we define the degree of orthogonality of a baseline $A_i$ as shown in Equation \ref{eqn:orthogonality}. $d(A_i)$ represents the total number of bug reports for which $A_i$ was able to assign correct developer or components while other baselines misclassified these bug reports.

\begin{equation}
\label{eqn:orthogonality}
   d(A_i) = |A_{T_i} - \left( \bigcup_{j\neq i}^n A_{T_j} \right)|
\end{equation}
In the above equation, $T$ denotes the set of bug reports used for evaluating all baselines, and $A_{T_i}$ denotes the set of bug reports which are correctly triaged by the baseline $A_i$. $\bigcup_{j\neq i}^n A_{T_j}$ represents the set of bug reports which are correctly triaged by other baselines except $A_i$. Thus, $d(A_i)$ is the cardinality of the set difference between $A_{T_i}$ and $\bigcup_{j\neq i}^n A_{T_j}$.

\subsection{\textbf{RQ}$_4$: Further Investigating Model Performance}

We aim to comprehend the nature of misclassified bug reports that are common among the baselines. Therefore, we hypothesize that the misclassification may occur due to textual similarity between the assigned developer or component. Suppose that we have a total of $m$ misclassified bug reports by all baselines, and let $C_1, C_2, \dots, C_n$ denote the actual classes of these reports. To find the similarity between the actual classes, we define $CP_i$ as the set of all bug reports that belong to class $C_i$ as follows:

\begin{equation}
CP_i = b_{i1} \oplus b_{i2} \oplus \dots \oplus b_{ij}
\end{equation}

Then, we transform $CP_i$ into a vector $CP^v_i$ using the TF-IDF method. The textual similarity between two classes ($C_i, C_{i^\prime}$) can be calculated as follows:

\begin{equation}
cosine\_similarity (C_i, C_{i^\prime})= \frac{CP^v_i \cdot CP^v_{i^\prime}}{|CP^v_i| |CP^v_{i^\prime}|}
\label{eq:cosine-similarity}
\end{equation}

Finally, we compute the correlation between the cosine similarity of ($C_i, C_{i^\prime}$) and the number of times $C_i$ is misclassified as $C_{i^\prime}$ or vice versa across all baselines. If there is a positive correlation, we can conclude that similarity is the reason behind the misclassification across baselines.

\section{Experimental Results and Analysis}\label{experiment}
For each research question (RQ), we present and discuss our findings in the following sections.

\subsection{\textbf{RQ}$_1$: Effectivness of the baselines in developer assignment}
\begin{figure*}[tb]
  \centering
  \begin{subfigure}[b]{0.45\linewidth}
    \includegraphics[width=\linewidth]{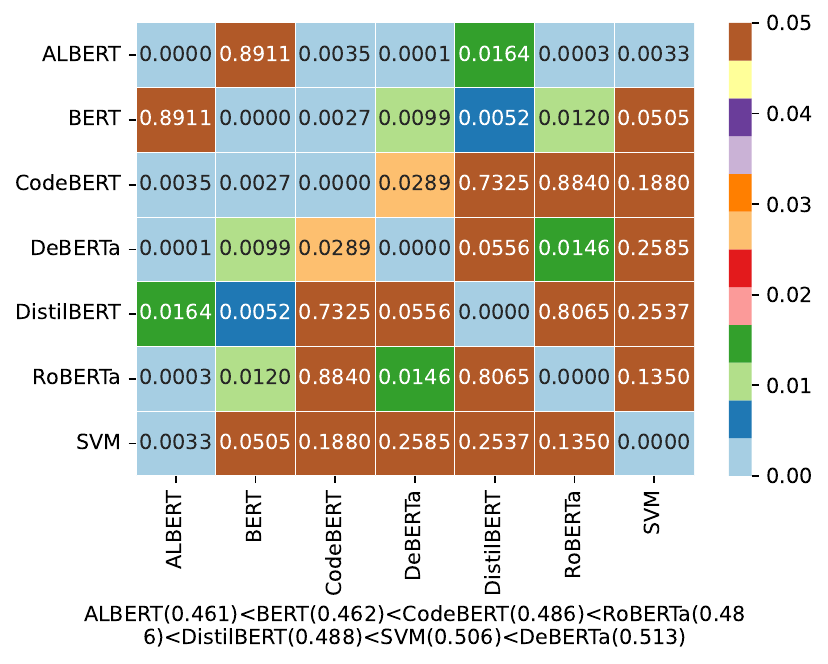}
    \caption{Eclipse JDT}
    \label{fig:ss-eclipse_jdt_developer}
  \end{subfigure}
  \hfill
  \begin{subfigure}[b]{0.45\linewidth}
    \includegraphics[width=\linewidth]{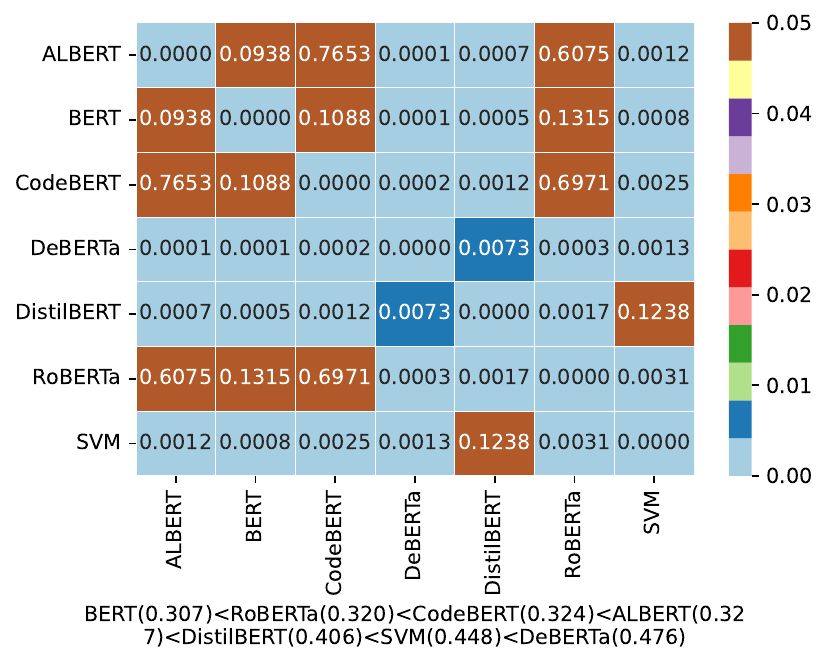}
    \caption{Eclipse Platform}
    \label{fig:ss-eclipse_platform_developer}
  \end{subfigure}

\medskip

\begin{subfigure}[b]{0.45\linewidth}
    \includegraphics[width=\linewidth]{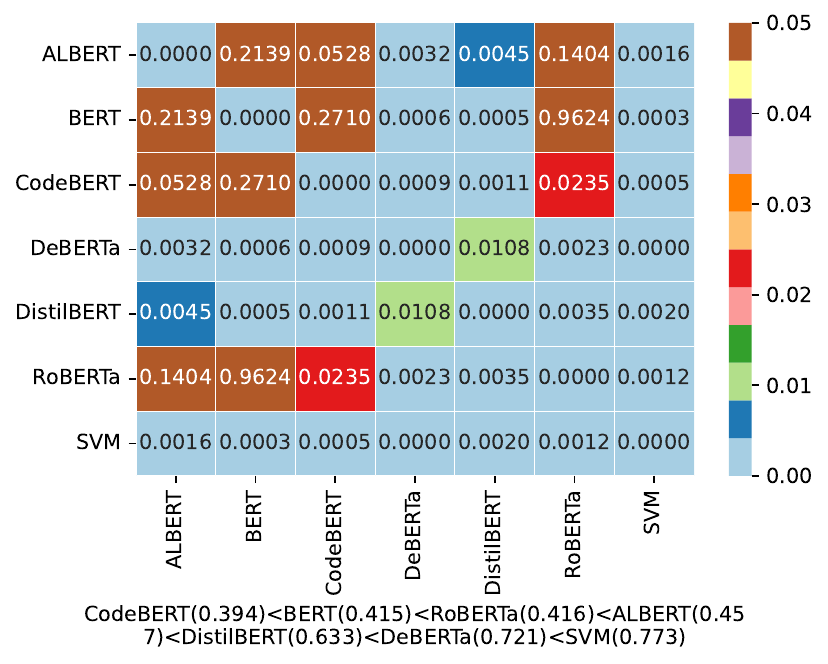}
    \caption{Mozilla Core}
    \label{fig:ss-mozilla_core_developer}
  \end{subfigure}
  \hfill
  \begin{subfigure}[b]{0.45\linewidth}
    \includegraphics[width=\linewidth]{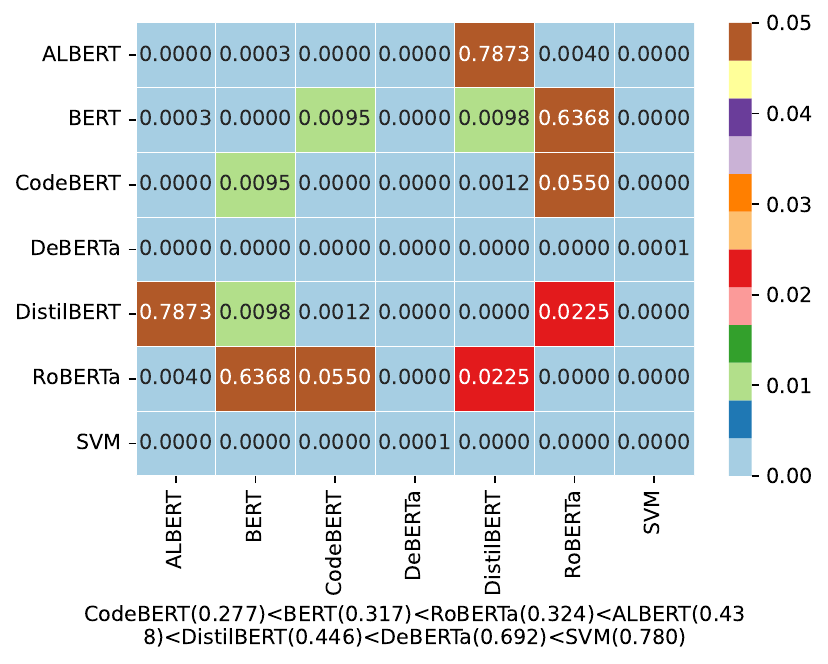}
    \caption{Mozilla SeaMonkey}
    \label{fig:ss-mozilla_seamoney_developer}
  \end{subfigure}
  
  \caption{Statistical significance test of the MMR scores of our baselines in terms of developer assignment across all datasets}
  \label{fig:ss-dev-assignment}
\end{figure*}

The primary aim of this RQ is to examine the effectiveness of our baselines in assigning developers to bug reports. To address this research question, we assess each baseline's performance using Top@1, Top@5, Top@10, and MRR metrics, which indicate how well the baseline ranks the correct developer as the top recommendation (Top@1), within the top 5 (Top@5), within the top 10 (Top@10), and on average across all the ranked lists (MRR). The average results for these evaluation metrics across all datasets are presented in Table \ref{tbl:results-dev-assignment}. Additionally, we determine whether the performance differences between baselines are statistically significant using their MRR scores. That is, we set $H_0$ as the average performance of our baselines being the same. The results of the statistical significance test between baselines in all datasets are shown in Fig. \ref{fig:ss-dev-assignment}. Each cell of Fig. \ref{fig:ss-eclipse_jdt_developer}, \ref{fig:ss-eclipse_platform_developer}, \ref{fig:ss-mozilla_core_developer}, and \ref{fig:ss-mozilla_seamoney_developer} represents the corresponding p-value of the calculated t-statistic (Equation \ref{t-stat}) between two baselines. For this statistical significance test, the significance level ($\alpha$) is set to 0.05, which means that if the p-value between two models is less than $\alpha$, $H_0$ can be rejected with a 95\% confidence level.

In Eclipse JDT dataset, DeBERTa achieves the highest average Top@1 and Top@5 scores as shown in Table \ref{tbl:results-dev-assignment}. Its Top@10 (0.866) score is almost equivalent to the SVM (0.873). Using the scores of MRR from Table \ref{tbl:results-dev-assignment}, the baselines can be ranked as ALBERT $<$ BERT $<$ CodeBERT $<$ RoBERTa $<$ DistilBERT $<$ SVM $<$ DeBERTa. However, the result of the statistical significance test in Fig. \ref{fig:ss-eclipse_jdt_developer} shows the performance difference between the top baselines (DistilBERT, SVM and DeBERTa) is not statistically significant. Other notable findings from Fig. \ref{fig:ss-eclipse_jdt_developer} is given below.

\begin{itemize}
    \item ALBERT's performance is not significantly better than any other model, but it is equivalent to BERT's performance
    \item BERT's performance is not significantly better than any other model, but it is equivalent to SVM and ALBERT's performance
    \item CodeBERT performs significantly better than ALBERT and BERT, and performs equally as well as DistilBERT, RoBERTa, and SVM
    \item DeBERTa performs significantly better than ALBERT, BERT, CodeBERT, and RoBERTa, but not better than SVM and DistilBERT
    \item DistilBERT performs significantly better than ALBERT and BERT, and performs equally as well as CodeBERT, RoBERTa, DeBERTa, and SVM
    \item RoBERTa performs significantly better than ALBERT and BERT, and performs equally as well as CodeBERT, DistilBERT, and SVM
    \item SVM performs significantly better than ALBERT only and performs equally as well as other baselines
\end{itemize}

In Eclipse Platform dataset, DeBERTa achieves the highest average scores in all three metrics. The ranking of baselines based on MRR is BERT $<$ RoBERTa $<$ CodeBERT $<$ ALBERT $<$ DistilBERT $<$ SVM $<$ DeBERTa. In this dataset, the top baseline's (DeBERTa) performance is statistically significant compared to others as displayed in \ref{fig:ss-eclipse_platform_developer}.

In both Mozilla datasets, SVM achieves the highest average scores in all three metrics. The ranking of baselines based on MRR is also consistent which is CodeBERT $<$ BERT $<$ RoBERTa $<$ ALBERT $<$ DistilBERT $<$ DeBERTa $<$ SVM. Here, SVM's performance is statistically significant compared to others as displayed in the Fig. \ref{fig:ss-mozilla_core_developer} \& \ref{fig:ss-mozilla_seamoney_developer}.

\textbf{Summary of RQ}$_1$: The results indicate that DeBERTa and SVM exhibit the best performance overall, with the highest average scores for Top@1, Top@5, Top@10, and MRR across multiple datasets. Notably, the performance of the models varies across datasets, with some models showing better results on specific datasets compared to others. Specifically, DeBERTa demonstrates statistically significant performance among all transformer-based neural text embedding techniques, except in the Eclipse JDT dataset, while SVM exhibits statistically significant performance among all baselines in both Mozilla~datasets.

\begin{table}[tb]
\centering
\setlength{\tabcolsep}{4pt}
\caption{Average Top@1, Top@5, Top@10 and MRR of Developer Assignment. The best-performing model for each dataset is highlighted in bold. }
\label{tbl:results-dev-assignment}
\begin{tabular}{l|l|l|l|l|l}
\hline
\textbf{Dataset}                   & \textbf{Model} & \textbf{Top@1}  & \textbf{Top@5}  & \textbf{Top@10} & \textbf{MRR}   \\ \hline
\multirow{7}{*}{Eclipse JDT}       & ALBERT         & 0.291          & 0.673          & 0.805          & 0.461          \\ \cline{2-6} 
                                   & BERT           & 0.298          & 0.663          & 0.791          & 0.462          \\ \cline{2-6} 
                                   & CodeBERT       & 0.323          & 0.696          & 0.812          & 0.486          \\ \cline{2-6} 
                                   & DeBERTa        & \textbf{0.344} & \textbf{0.731} & 0.866          & \textbf{0.513} \\ \cline{2-6} 
                                   & DistilBERT     & 0.319          & 0.703          & 0.838          & 0.487          \\ \cline{2-6} 
                                   & RoBERTa        & 0.321          & 0.695          & 0.816          & 0.486          \\ \cline{2-6} 
                                   & SVM            & 0.333          & 0.727          & \textbf{0.873} & 0.506          \\ \hline
\multirow{7}{*}{Eclipse Platform}  & ALBERT         & 0.203          & 0.458          & 0.570          & 0.327          \\ \cline{2-6} 
                                   & BERT           & 0.191          & 0.428          & 0.539          & 0.307          \\ \cline{2-6} 
                                   & CodeBERT       & 0.199          & 0.460          & 0.572          & 0.323          \\ \cline{2-6} 
                                   & DeBERTa        & \textbf{0.317} & \textbf{0.672} & \textbf{0.795} & \textbf{0.476} \\ \cline{2-6} 
                                   & DistilBERT     & 0.262          & 0.579          & 0.697          & 0.406          \\ \cline{2-6} 
                                   & RoBERTa        & 0.195          & 0.456          & 0.570          & 0.320          \\ \cline{2-6} 
                                   & SVM            & 0.289          & 0.643          & 0.771          & 0.448          \\ \hline
\multirow{7}{*}{Mozilla Core}      & ALBERT         & 0.392          & 0.509          & 0.569          & 0.457          \\ \cline{2-6} 
                                   & BERT           & 0.346          & 0.467          & 0.536          & 0.415          \\ \cline{2-6} 
                                   & CodeBERT       & 0.331          & 0.441          & 0.506          & 0.394          \\ \cline{2-6} 
                                   & DeBERTa        & 0.653          & 0.795          & 0.838          & 0.721          \\ \cline{2-6} 
                                   & DistilBERT     & 0.562          & 0.709          & 0.759          & 0.633          \\ \cline{2-6} 
                                   & RoBERTa        & 0.354          & 0.465          & 0.525          & 0.416          \\ \cline{2-6} 
                                   & SVM            & \textbf{0.676} & \textbf{0.895} & \textbf{0.923} & \textbf{0.773} \\ \hline
\multirow{7}{*}{Mozilla SeaMonkey} & ALBERT         & 0.351          & 0.509          & 0.605          & 0.438          \\ \cline{2-6} 
                                   & BERT           & 0.219          & 0.387          & 0.511          & 0.317          \\ \cline{2-6} 
                                   & CodeBERT       & 0.180          & 0.354          & 0.466          & 0.277          \\ \cline{2-6} 
                                   & DeBERTa        & 0.613          & 0.778          & 0.845          & 0.692          \\ \cline{2-6} 
                                   & DistilBERT     & 0.347          & 0.535          & 0.625          & 0.446          \\ \cline{2-6} 
                                   & RoBERTa        & 0.229          & 0.408          & 0.517          & 0.324          \\ \cline{2-6} 
                                   & SVM            & \textbf{0.682} & \textbf{0.903} & \textbf{0.944} & \textbf{0.780} \\ \hline
\end{tabular}
\end{table}

\subsection{\textbf{RQ}$_2$: Effectiveness of baselines in component assignment}
\begin{figure*}[tb]
  \centering
  \begin{subfigure}[b]{0.45\linewidth}
    \includegraphics[width=\linewidth]{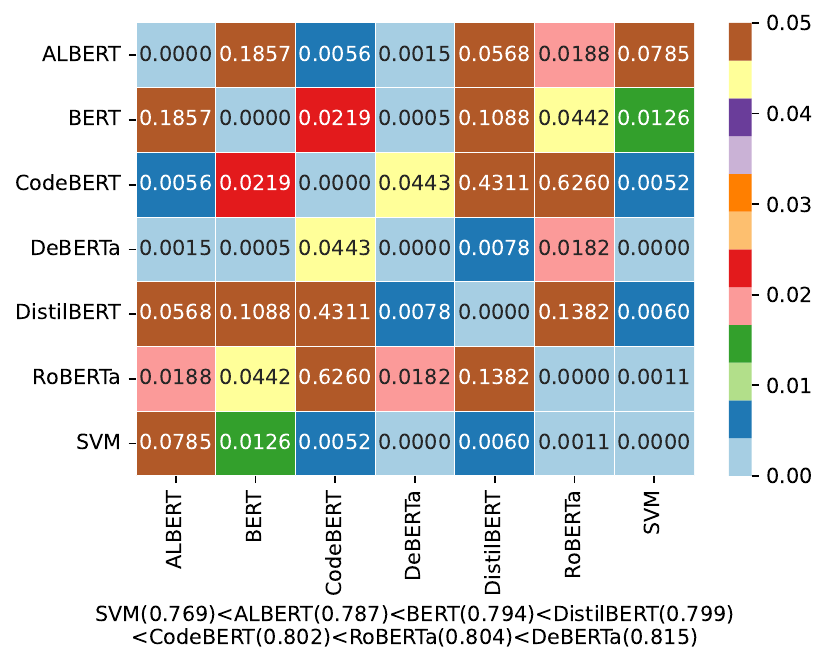}
    \caption{Eclipse JDT}
    \label{fig:ss-eclipse_jdt_component}
  \end{subfigure}
  \hfill
  \begin{subfigure}[b]{0.45\linewidth}
    \includegraphics[width=\linewidth]{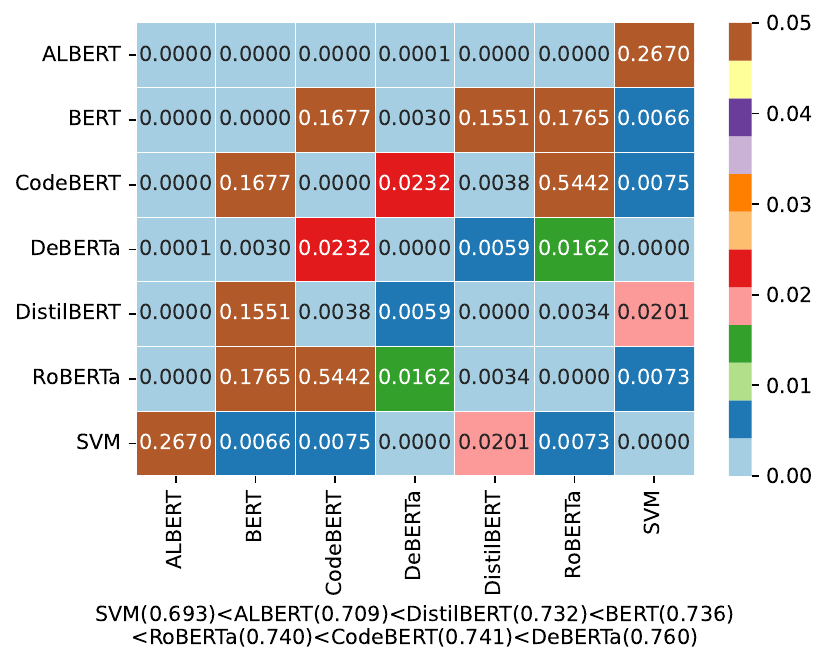}
    \caption{Eclipse Platform}
    \label{fig:ss-eclipse_platform_component}
  \end{subfigure}

\medskip

\begin{subfigure}[b]{0.45\linewidth}
    \includegraphics[width=\linewidth]{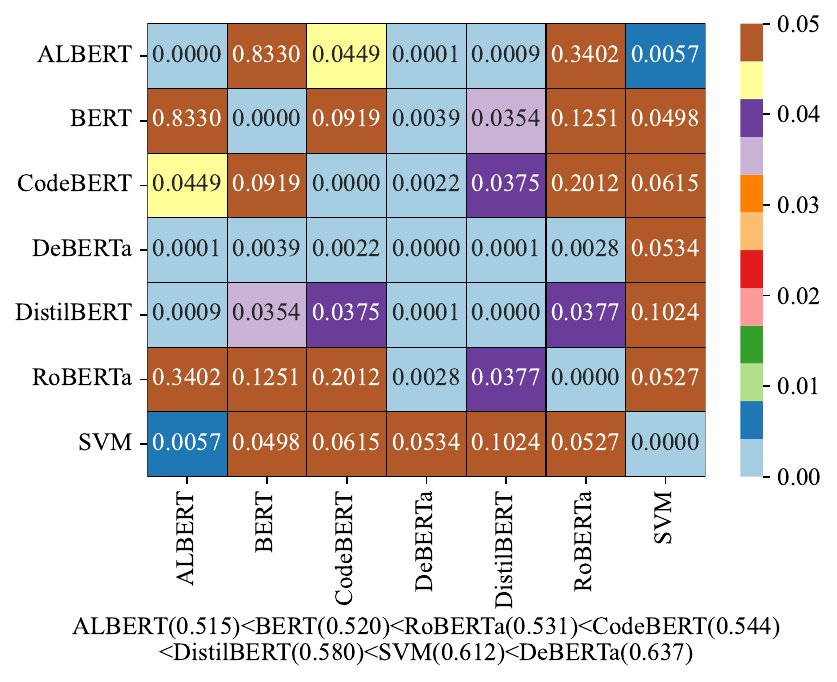}
    \caption{Mozilla Core}
    \label{fig:ss-mozilla_core_component}
  \end{subfigure}
  \hfill
  \begin{subfigure}[b]{0.45\linewidth}
    \includegraphics[width=\linewidth]{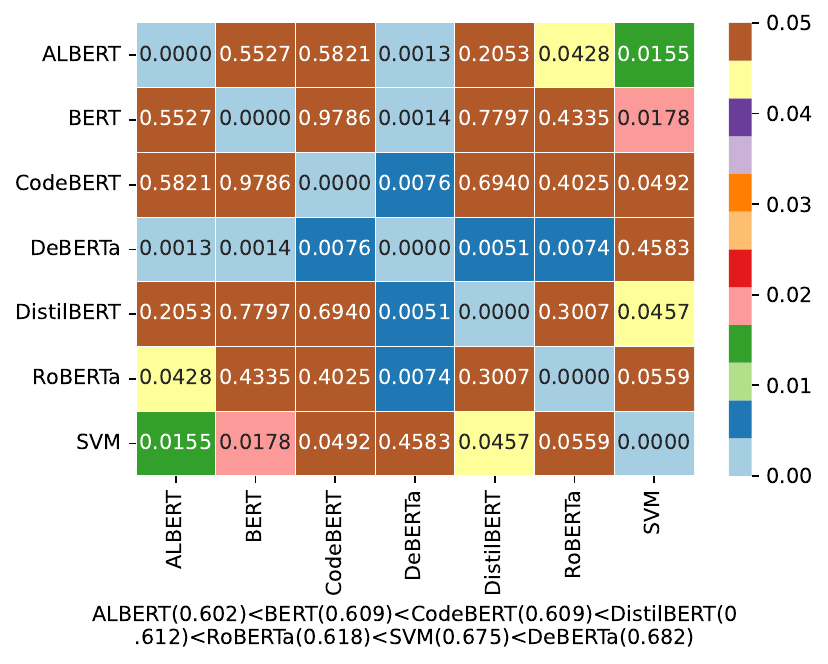}
    \caption{Mozilla SeaMonkey}
    \label{fig:ss-mozilla_seamoney_component}
  \end{subfigure}
  
  \caption{Statistical significance test of F1-Scores of our baselines in terms of component assignment across all datasets}
  \label{fig:ss-comp-assignment}
\end{figure*}

The aim of this RQ is to evaluate the effectiveness of the baselines in assigning components to bug reports. Precision, Recall, and F1-Score are used as evaluation metrics, and the average results for these metrics across all datasets are presented in Table \ref{tbl:results-comp-assignment}. Similar to the developer assignment task, we conduct a statistical significance test between the baselines based on F1-Scores, using the same $\alpha$ and $H_0$. The results of this test for all datasets are shown in Fig. \ref{fig:ss-comp-assignment}.

DeBERTa achieves the highest average Precision in Eclipse JDT and Mozilla SeaMonkey datasets, while CodeBERT and BERT achieve the highest average Precision in Eclipse Platform and Mozilla Core datasets, respectively. In terms of Recall and F1-score, DeBERTa consistently obtains the highest average scores across all datasets, as demonstrated in Table \ref{tbl:results-comp-assignment}.

However, while F1-scores indicate that DeBERTa is the best performer across all datasets, the statistical significance test reveals that its performance is similar to SVM in both Mozilla datasets, as depicted in Fig. \ref{fig:ss-mozilla_core_component} \& \ref{fig:ss-mozilla_seamoney_component}. Nevertheless, in both Eclipse datasets, DeBERTa's performance is statistically significant compared to other baselines, as shown in Fig.~\ref{fig:ss-eclipse_jdt_component}~\&~\ref{fig:ss-eclipse_platform_component}.

\textbf{Summary of RQ}$_2$: Unlike the developer assignment task, the performance of the models does not vary significantly across datasets. From Table \ref{tbl:results-comp-assignment} and Fig. \ref{fig:ss-comp-assignment}, it is evident that DeBERTa performs consistently and significantly better than other models in assigning correct components to bug reports.

\begin{table}[tb]
\centering
\setlength{\tabcolsep}{4pt}
\caption{Average Precision, Recall and F1-Score of Component Assignment. The best-performing model for each dataset is highlighted in bold.}
\label{tbl:results-comp-assignment}
\begin{tabular}{l|l|l|l|l}
\hline
\textbf{Dataset}                            & \textbf{Model}      & \textbf{Precision}      & \textbf{Recall}         & \textbf{F1-Score}       \\ \hline
\multirow{7}{*}{Eclipse JDT}       & ALBERT     & 0.808          & 0.798          & 0.787          \\ \cline{2-5} 
                                   & BERT       & 0.814          & 0.803          & 0.794          \\ \cline{2-5} 
                                   & CodeBERT   & 0.823          & 0.813          & 0.802          \\ \cline{2-5} 
                                   & DeBERTa    & \textbf{0.826} & \textbf{0.822} & \textbf{0.815} \\ \cline{2-5} 
                                   & DistilBERT & 0.817          & 0.809          & 0.799          \\ \cline{2-5} 
                                   & RoBERTa    & 0.823          & 0.814          & 0.804          \\ \cline{2-5} 
                                   & SVM        & 0.783          & 0.777          & 0.769          \\ \hline
\multirow{7}{*}{Eclipse Platform}  & ALBERT     & 0.748          & 0.724          & 0.709          \\ \cline{2-5} 
                                   & BERT       & 0.771          & 0.751          & 0.736          \\ \cline{2-5} 
                                   & CodeBERT   & \textbf{0.784} & 0.755          & 0.741          \\ \cline{2-5} 
                                   & DeBERTa    & 0.780          & \textbf{0.767} & \textbf{0.760} \\ \cline{2-5} 
                                   & DistilBERT & 0.768          & 0.748          & 0.732          \\ \cline{2-5} 
                                   & RoBERTa    & 0.780          & 0.755          & 0.740          \\ \cline{2-5} 
                                   & SVM        & 0.721          & 0.702          & 0.693          \\ \hline
\multirow{7}{*}{Mozilla Core}      & ALBERT     & 0.646          & 0.546          & 0.515          \\ \cline{2-5} 
                                   & BERT       & \textbf{0.684} & 0.560          & 0.520          \\ \cline{2-5} 
                                   & CodeBERT   & 0.683          & 0.581          & 0.544          \\ \cline{2-5} 
                                   & DeBERTa    & 0.678          & \textbf{0.644} & \textbf{0.637} \\ \cline{2-5} 
                                   & DistilBERT & 0.669          & 0.605          & 0.580          \\ \cline{2-5} 
                                   & RoBERTa    & 0.680          & 0.570          & 0.531          \\ \cline{2-5} 
                                   & SVM        & 0.664          & 0.618          & 0.612          \\ \hline
\multirow{7}{*}{Mozilla SeaMonkey} & ALBERT     & 0.716          & 0.626          & 0.602          \\ \cline{2-5} 
                                   & BERT       & 0.713          & 0.640          & 0.609          \\ \cline{2-5} 
                                   & CodeBERT   & 0.719          & 0.642          & 0.609          \\ \cline{2-5} 
                                   & DeBERTa    & \textbf{0.741} & \textbf{0.696} & \textbf{0.682} \\ \cline{2-5} 
                                   & DistilBERT & 0.721          & 0.642          & 0.612          \\ \cline{2-5} 
                                   & RoBERTa    & 0.717          & 0.650          & 0.618          \\ \cline{2-5} 
                                   & SVM        & 0.737          & 0.683          & 0.675          \\ \hline
\end{tabular}
\end{table}

\subsection{\textbf{RQ}$_3$: Degree of orthogonality between baselines}
In RQ$_1$ and RQ$2$, it was discussed that certain baselines show similar performance in terms of both developer and component assignment. This RQ aims to investigate if these baselines perform better on an independent or the same set of bug reports. To illustrate the motivation behind this RQ, a toy example is presented where two baselines, $A_1$ and $A_2$, are tested on a set of bug reports $T = \{B_1, B_2, \dots, B{10}\}$. When statistical significance indicates that $A_1$ and $A_2$ perform similarly, there could be several cases.

The first case is when both baselines correctly identify the same set of bug reports, such as $\{B_2, B_{3}, B_4\}$. The second case is when both baselines correctly identify two different sets of bug reports, such as $A_1 = \{B_2, B_3, B_4\}$ and $A_2 = \{B_4, B_5, B_6\}$. The third case is when there are some common bug reports between the correct predictions, such as $A_1 = \{B_2, B_3, B_4\}$ and $A_2 = \{B_2, B_5, B_6\}$.

\begin{figure*}[tb]
  \centering
  \begin{subfigure}[b]{0.45\linewidth}
    \includegraphics[width=\linewidth]{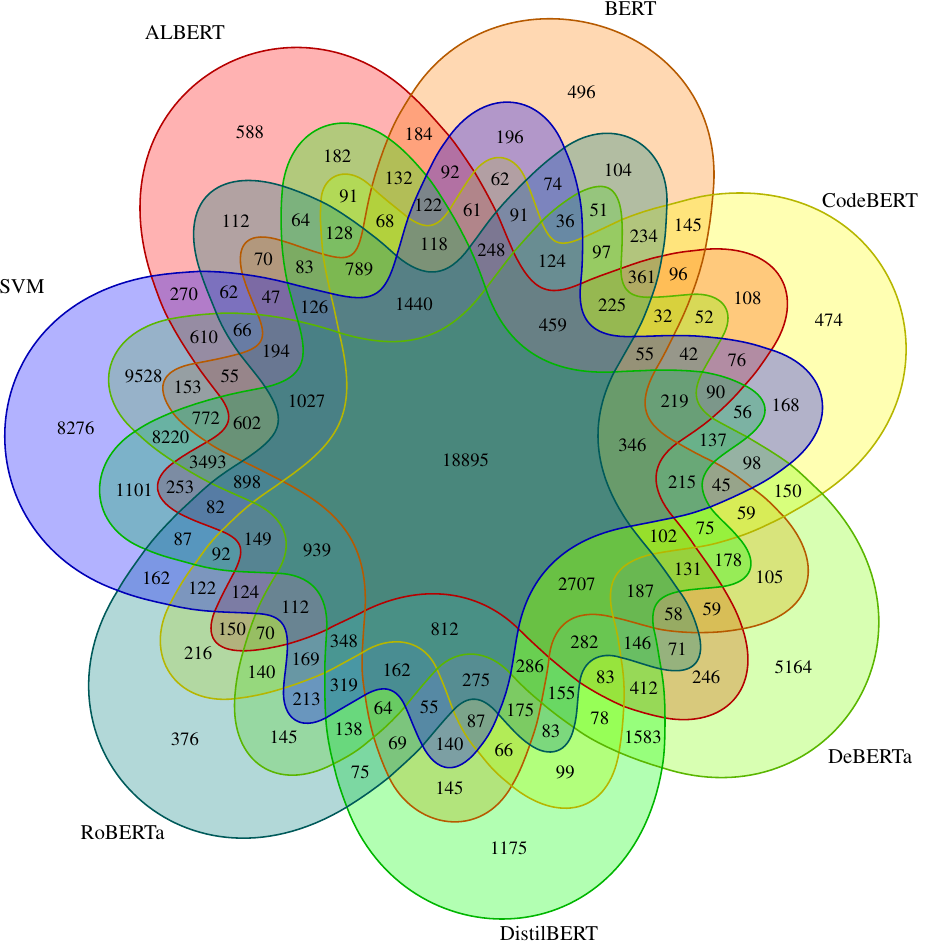}
    \caption{Degree of orthogonality in Developer Assignment}
    \label{fig:correct_developer_orthogonality}
  \end{subfigure}
  \hfill
  \begin{subfigure}[b]{0.45\linewidth}
    \includegraphics[width=\linewidth]{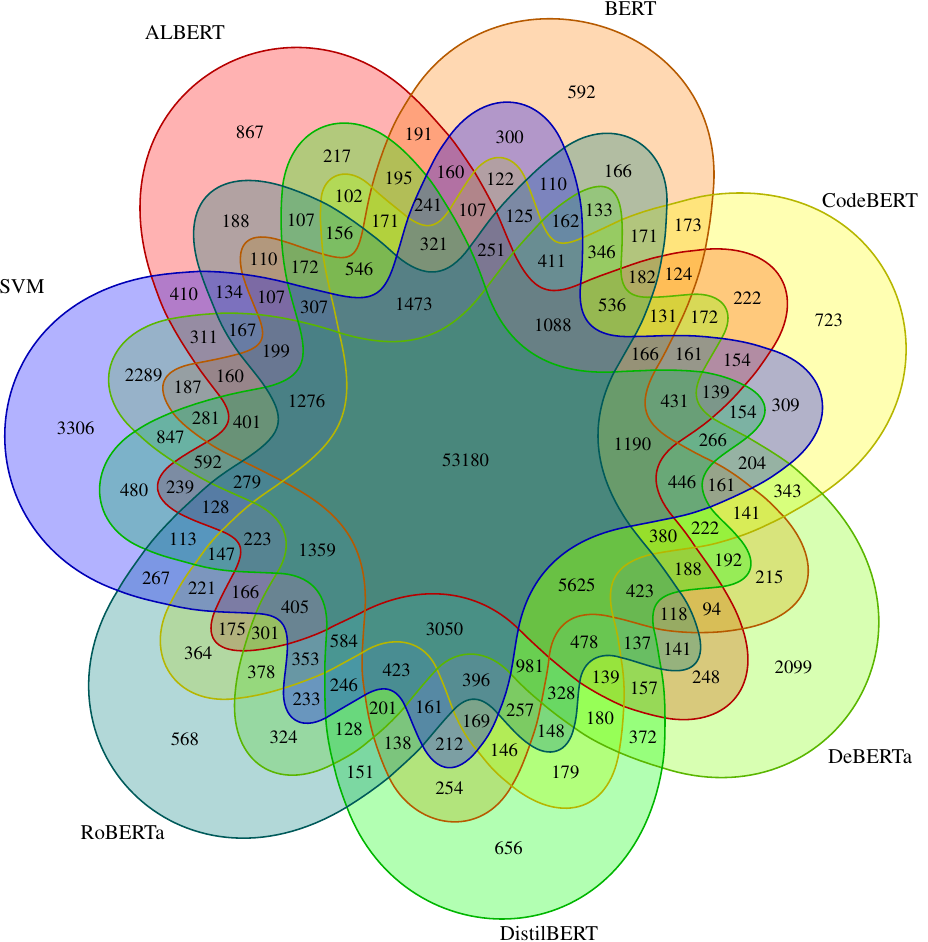}
    \caption{Degree of orthogonality in Component Assignment}
    \label{fig:correct_component_orthogonality}
  \end{subfigure}
  \caption{Degree of orthogonality based on correct predictions of all baselines across all dataset in term of developer and component assignment}
  \label{fig:degree-of-otho-correct-dev-comp}
\end{figure*}

All cases indicate that both baselines are similar based on MRR or F1-score and statistical significance tests. However, if the second or third case occurs, it suggests that there is uniqueness between the baselines as they perform better for two different sets of bug reports, indicating orthogonality between $A_1$ and $A_2$. Fig. \ref{fig:degree-of-otho-correct-dev-comp} presents the result of the orthogonality analysis across datasets and baselines based on correct predictions. Top@1 accuracy is used to measure the degree of orthogonality in developer assignment, while precision is used to calculate the degree of orthogonality in component assignment.

In Fig. \ref{fig:correct_developer_orthogonality} or \ref{fig:correct_component_orthogonality}, the value in the common area indicates the cardinality of the set of bug reports from all datasets that are correctly assigned to developers or components, respectively, by all baselines. For instance, in Fig. \ref{fig:correct_developer_orthogonality}, all baselines assign 18895 bugs to the correct developer from all datasets. Conversely, in Fig. \ref{fig:correct_component_orthogonality}, all baselines assign 53180 bugs to the correct component from all datasets. The exclusive areas in Fig. \ref{fig:correct_developer_orthogonality} and \ref{fig:correct_component_orthogonality} indicate the degree of orthogonality of each baseline. For example, ALBERT has a degree of orthogonality of 588 (Fig. \ref{fig:correct_developer_orthogonality}) in the developer assignment task, whereas its degree of orthogonality is 867 (Fig. \ref{fig:correct_component_orthogonality}) in the component assignment task. Similar to ALBERT, other baselines have also a significant degree of orthogonality in both task. Additionally, the degree of commonality between baselines can be found in Fig. \ref{fig:correct_developer_orthogonality} and \ref{fig:correct_component_orthogonality}. For instance, in Fig. \ref{fig:correct_developer_orthogonality}, SVM and DeBERTa together assign correct developers to 9528 bug reports.

\textbf{Summary of RQ$_3$}: In conclusion, RQ$_3$ reveals that there is orthogonality or uniqueness between baselines in both developer and component assignment tasks, despite their similarities in MRR, F1-score or statistical significance test results. Among all baselines, SVM demonstrates the highest degree of orthogonality in both tasks. Meanwhile, among all transformer-based techniques, DeBERTa displays the highest degree of orthogonality.

\subsection{\textbf{RQ}$_4$: Why do baselines fail or possess unique behavior?}
In this RQ, we try to understand when all baselines misclasify a set of bug reports, and when they exhibit unique behavior.

\begin{figure}[tb]
    \centering
    \includegraphics[scale=0.50]{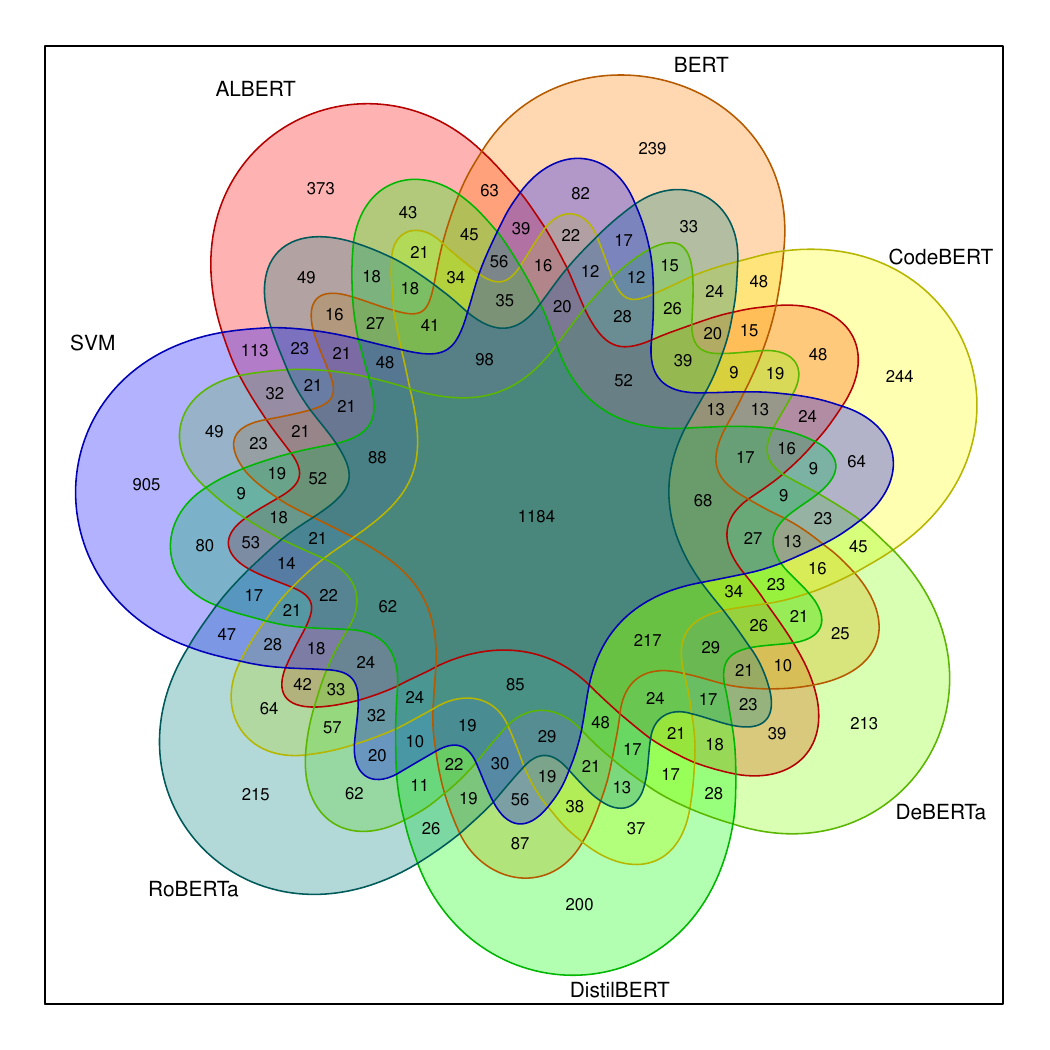}
    \caption{Degree of orthogonality of based on incorrect component assignments in Eclipse JDT dataset}
    \label{fig:degree-of-ortho-comp-mis-classification}
\end{figure}

\begin{figure*}[tb]
 \centering
 \begin{subfigure}[b]{0.45\linewidth}
   \includegraphics[width=\linewidth]{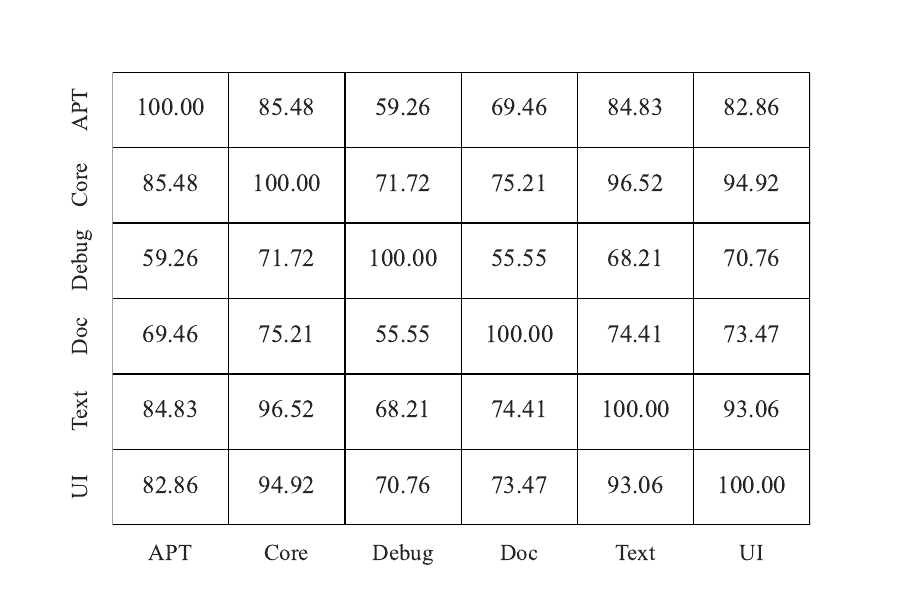}
   \caption{Similarity between components}
   \label{fig:in_correct_sim_eclipse_jdt}
 \end{subfigure}
 \hfill
 \begin{subfigure}[b]{0.45\linewidth}
   \includegraphics[width=\linewidth]{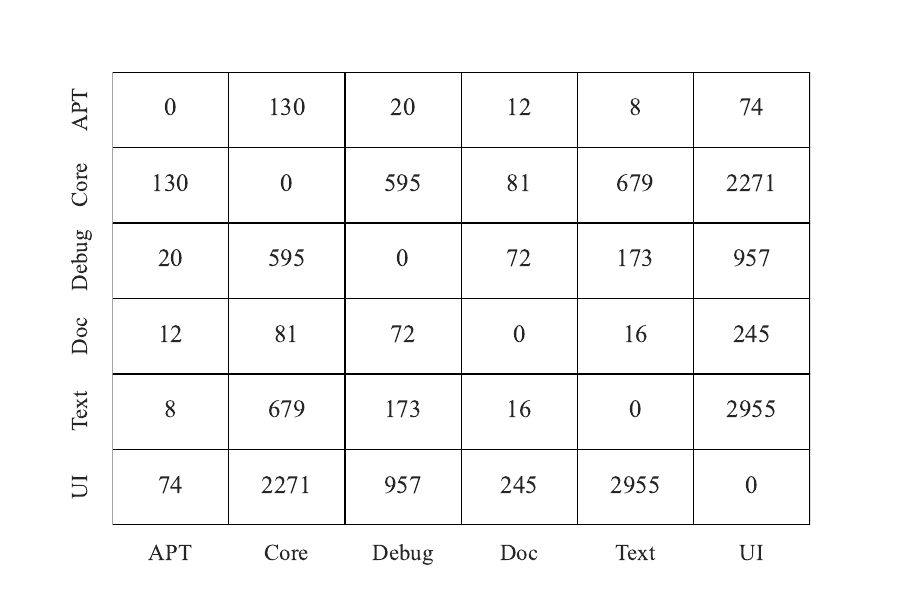}
   \caption{Confusion between components}
   \label{fig:in_correct_overlap_eclipse_jdt}
 \end{subfigure}
 \caption{Similarity and confusion between the components in Eclipse JDT}
 \label{fig:similarity-vs-confusion}
\end{figure*}

Fig. \ref{fig:degree-of-ortho-comp-mis-classification} presents the result of the orthogonality analysis based on incorrect component assignments in Eclipse JDT. In this figure, the exclusive areas represent each model's degree of orthogonality of misclassification. The value in the common area indicates the cardinality of the set of bug reports misclassified by all baselines. That is, all baselines fail to assign correct component to 1184 bug reports. Fig. \ref{fig:in_correct_sim_eclipse_jdt} represents the cosine similarity (Eqn. \ref{eq:cosine-similarity}) between the actual components of these bug reports. For instance, UI \& Text are 93.06\% similar, and UI \& Core are 94.92\% similar. Fig. \ref{fig:in_correct_sim_eclipse_jdt} represents the confusion matrix of 1184 bug reports across all baselines. Each cell of this figure represents the total number of confusions across all baselines. For example, in total 2955 times UI was assigned as Text or vice-versa by all baselines. Similarly, in total 2271 times UI was assigned as Core or vice-versa by all baselines. From Fig. \ref{fig:in_correct_sim_eclipse_jdt} \& \ref{fig:in_correct_sim_eclipse_jdt}, it is apparent that when the similarity between two components is high, baselines fail to assign the correct component. To understand it better, we measure correlation between the similarity and the total number of confusions of all baselines. Our result indicates that there is positive correlation between these two. We conduct the same study for both developer and component assignment in other datasets and found positive correlation also. Since the number of components and developers in other datasets is very high (Table \ref{tbl:dataset}), we could not present and discuss it here.

\textbf{Summary of RQ$_4$}: Similarity between components and developers is a big issue for all baselines. When there is a high textual similarity between components or developers, no baseline can identify them correctly.

\section{Threats to Validity}
In this section, we discuss some potential validity threats of our work which we want to focus in our future research.

\textbf{External Validity:} Our study focuses solely on experiments conducted using transformer-based deep learning techniques. While we did not compare these techniques to other deep learning methods such as LSTM and CNN, it's worth noting that transformer-based neural text embedding techniques have been shown to outperform other deep learning-based techniques in various NLP tasks, including text summarization, question-answering, and document classification \cite{vaswani2017attention,devlin2018bert}. Additionally, the artifacts from existing works on bug triaging \cite{lee2022light,zhang2020efficient} are not publicly available, which makes it challenging to compare transformer-based techniques with other deep learning-based techniques. As a result, our study may be limited in its external validity.

\textbf{Conclusion Validity:} Based on the findings we gathered from four open-source datasets, we make all statistical conclusions regarding the baselines. However, it's important to note that these statistical conclusions may or may not apply to industrial projects. As there is currently no publicly available industrial dataset, our plan for the future is to collect datasets from the industry and conduct a similar study.

\section{Conclusions \& Future Work}
\label{conclusion}

We have presented a comprehensive empirical study that examines the potential for using pre-trained transformer language models, trained on both code and natural language across a variety of architectures, to automate the task of automated bug triaging. Our study context consists of 136k bug reports collected from a combined 53 years of maintenance history from four popular open source software repositories. These repositories consist of 165 distinct software components to which a combined 462 developers have made contributions. The results of our study illustrate that DeBERTa performs significantly better other transformer-based language models in both developer and component assignment tasks. Somewhat surprisingly, the simpler TF-IDF-based SVM baseline performs best, for the task for developer assignment, on two of our four studied datasets, illustrating that a well-tuned term, frequency-based approach can perform well when triaging bugs to developers. Each technique has a certain degree of orthogonality which indicates that the unique properties of each technique allow them to capture a different set of patterns from bug reports. Similarity between components and developers hampers the performance of all transformer-based language models. There is a positive correlation between these similarities and number of misclassifications.

Some of the findings of our study illustrate promising avenues for future work. First, given that models seem to struggle when there is high textual similarity between developers and components, project maintainers should be aware of these situations and look toward other techniques for bug triaging. The research community should investigate alternative techniques for triaging when likely assigned developers are similar to one another. Second, given that there is a somewhat substantial degree of orthogonality between our studied techniques, future work should examine the potential for ensemble techniques that combine the relative strengths of each model such that overall performance might be improved.

\balance
\bibliographystyle{IEEEtran}
\bibliography{references}

\begin{thebibliography}{10}
\providecommand{\url}[1]{#1}
\csname url@samestyle\endcsname
\providecommand{\newblock}{\relax}
\providecommand{\bibinfo}[2]{#2}
\providecommand{\BIBentrySTDinterwordspacing}{\spaceskip=0pt\relax}
\providecommand{\BIBentryALTinterwordstretchfactor}{4}
\providecommand{\BIBentryALTinterwordspacing}{\spaceskip=\fontdimen2\font plus
\BIBentryALTinterwordstretchfactor\fontdimen3\font minus
  \fontdimen4\font\relax}
\providecommand{\BIBforeignlanguage}[2]{{%
\expandafter\ifx\csname l@#1\endcsname\relax
\typeout{** WARNING: IEEEtran.bst: No hyphenation pattern has been}%
\typeout{** loaded for the language `#1'. Using the pattern for}%
\typeout{** the default language instead.}%
\else
\language=\csname l@#1\endcsname
\fi
#2}}
\providecommand{\BIBdecl}{\relax}
\BIBdecl

\bibitem{anvik2006should}
J.~Anvik, L.~Hiew, and G.~C. Murphy, ``Who should fix this bug?'' in
  \emph{Proceedings of the 28th international conference on Software
  engineering}, 2006, pp. 361--370.

\bibitem{anvik2006automating}
J.~Anvik, ``Automating bug report assignment,'' in \emph{Proceedings of the
  28th international conference on Software engineering}, 2006, pp. 937--940.

\bibitem{crowston2006information}
K.~Crowston, J.~Howison, and H.~Annabi, ``Information systems success in free
  and open source software development: Theory and measures,'' \emph{Software
  Process: Improvement and Practice}, vol.~11, no.~2, pp. 123--148, 2006.

\bibitem{jeong2009improving}
G.~Jeong, S.~Kim, and T.~Zimmermann, ``Improving bug triage with bug tossing
  graphs,'' in \emph{Proceedings of the 7th joint meeting of the European
  software engineering conference and the ACM SIGSOFT symposium on The
  foundations of software engineering}, 2009, pp. 111--120.

\bibitem{zhang2016literature}
T.~Zhang, H.~Jiang, X.~Luo, and A.~T. Chan, ``A literature review of research
  in bug resolution: Tasks, challenges and future directions,'' \emph{The
  Computer Journal}, vol.~59, no.~5, pp. 741--773, 2016.

\bibitem{bhattacharya2012automated}
P.~Bhattacharya, I.~Neamtiu, and C.~R. Shelton, ``Automated, highly-accurate,
  bug assignment using machine learning and tossing graphs,'' \emph{Journal of
  Systems and Software}, vol.~85, no.~10, pp. 2275--2292, 2012.

\bibitem{kim2006long}
S.~Kim and E.~J. Whitehead~Jr, ``How long did it take to fix bugs?'' in
  \emph{Proceedings of the 2006 international workshop on Mining software
  repositories}, 2006, pp. 173--174.

\bibitem{xuan2017automatic}
J.~Xuan, H.~Jiang, Z.~Ren, J.~Yan, and Z.~Luo, ``Automatic bug triage using
  semi-supervised text classification,'' \emph{arXiv preprint
  arXiv:1704.04769}, 2017.

\bibitem{yang2014towards}
G.~Yang, T.~Zhang, and B.~Lee, ``Towards semi-automatic bug triage and severity
  prediction based on topic model and multi-feature of bug reports,'' in
  \emph{2014 IEEE 38th Annual Computer Software and Applications
  Conference}.\hskip 1em plus 0.5em minus 0.4em\relax IEEE, 2014, pp. 97--106.

\bibitem{lee2017applying}
S.-R. Lee, M.-J. Heo, C.-G. Lee, M.~Kim, and G.~Jeong, ``Applying deep learning
  based automatic bug triager to industrial projects,'' in \emph{Proceedings of
  the 2017 11th Joint Meeting on foundations of software engineering}, 2017,
  pp. 926--931.

\bibitem{mani2019deeptriage}
S.~Mani, A.~Sankaran, and R.~Aralikatte, ``Deeptriage: Exploring the
  effectiveness of deep learning for bug triaging,'' in \emph{Proceedings of
  the ACM India Joint International Conference on Data Science and Management
  of Data}, 2019, pp. 171--179.

\bibitem{church2017word2vec}
K.~W. Church, ``Word2vec,'' \emph{Natural Language Engineering}, vol.~23,
  no.~1, pp. 155--162, 2017.

\bibitem{PetersNIGCLZ18}
\BIBentryALTinterwordspacing
M.~E. Peters, M.~Neumann, M.~Iyyer, M.~Gardner, C.~Clark, K.~Lee, and
  L.~Zettlemoyer, ``Deep contextualized word representations,'' in
  \emph{Proceedings of the 2018 Conference of the North American Chapter of the
  Association for Computational Linguistics: Human Language Technologies,
  {NAACL-HLT} 2018, New Orleans, Louisiana, USA, June 1-6, 2018, Volume 1 (Long
  Papers)}, M.~A. Walker, H.~Ji, and A.~Stent, Eds.\hskip 1em plus 0.5em minus
  0.4em\relax Association for Computational Linguistics, 2018, pp. 2227--2237.
  [Online]. Available: \url{https://doi.org/10.18653/v1/n18-1202}
\BIBentrySTDinterwordspacing

\bibitem{zaidi2020applying}
S.~F.~A. Zaidi, F.~M. Awan, M.~Lee, H.~Woo, and C.-G. Lee, ``Applying
  convolutional neural networks with different word representation techniques
  to recommend bug fixers,'' \emph{IEEE Access}, vol.~8, pp.
  213\,729--213\,747, 2020.

\bibitem{devlin2018bert}
J.~Devlin, M.-W. Chang, K.~Lee, and K.~Toutanova, ``Bert: Pre-training of deep
  bidirectional transformers for language understanding,'' \emph{arXiv preprint
  arXiv:1810.04805}, 2018.

\bibitem{appendix}
Anonymous, ``Neural bug triaging online appendix,''
  \url{https://sagelab.io/neural-bug-triaging}, 2023.

\bibitem{xia2013accurate}
X.~Xia, D.~Lo, X.~Wang, and B.~Zhou, ``Accurate developer recommendation for
  bug resolution,'' in \emph{2013 20th Working Conference on Reverse
  Engineering (WCRE)}.\hskip 1em plus 0.5em minus 0.4em\relax IEEE, 2013, pp.
  72--81.

\bibitem{nath2021principal}
V.~Nath, D.~Sheldon, and J.~Alphonso-Gibbs, ``Principal component analysis and
  entropy-based selection for the improvement of bug triage,'' in \emph{2021
  20th IEEE International Conference on Machine Learning and Applications
  (ICMLA)}.\hskip 1em plus 0.5em minus 0.4em\relax IEEE, 2021, pp. 541--546.

\bibitem{nguyen2014topic}
T.~T. Nguyen, A.~T. Nguyen, and T.~N. Nguyen, ``Topic-based, time-aware bug
  assignment,'' \emph{ACM SIGSOFT Software Engineering Notes}, vol.~39, no.~1,
  pp. 1--4, 2014.

\bibitem{murphy2004automatic}
G.~Murphy and D.~Cubranic, ``Automatic bug triage using text categorization,''
  in \emph{Proceedings of the Sixteenth International Conference on Software
  Engineering \& Knowledge Engineering}.\hskip 1em plus 0.5em minus 0.4em\relax
  Citeseer, 2004, pp. 1--6.

\bibitem{anvik2011reducing}
J.~Anvik and G.~C. Murphy, ``Reducing the effort of bug report triage:
  Recommenders for development-oriented decisions,'' \emph{ACM Transactions on
  Software Engineering and Methodology (TOSEM)}, vol.~20, no.~3, pp. 1--35,
  2011.

\bibitem{sarkar2019improving}
A.~Sarkar, P.~C. Rigby, and B.~Bartalos, ``Improving bug triaging with high
  confidence predictions at ericsson,'' in \emph{2019 IEEE International
  Conference on Software Maintenance and Evolution (ICSME)}.\hskip 1em plus
  0.5em minus 0.4em\relax IEEE, 2019, pp. 81--91.

\bibitem{lin2009empirical}
Z.~Lin, F.~Shu, Y.~Yang, C.~Hu, and Q.~Wang, ``An empirical study on bug
  assignment automation using chinese bug data,'' in \emph{2009 3rd
  International Symposium on Empirical Software Engineering and
  Measurement}.\hskip 1em plus 0.5em minus 0.4em\relax IEEE, 2009, pp.
  451--455.

\bibitem{ahsan2009automatic}
S.~N. Ahsan, J.~Ferzund, and F.~Wotawa, ``Automatic software bug triage system
  (bts) based on latent semantic indexing and support vector machine,'' in
  \emph{2009 Fourth International Conference on Software Engineering
  Advances}.\hskip 1em plus 0.5em minus 0.4em\relax IEEE, 2009, pp. 216--221.

\bibitem{nasim2011automated}
S.~Nasim, S.~Razzaq, and J.~Ferzund, ``Automated change request triage using
  alpha frequency matrix,'' in \emph{2011 Frontiers of Information
  Technology}.\hskip 1em plus 0.5em minus 0.4em\relax IEEE, 2011, pp. 298--302.

\bibitem{florea2017spark}
A.-C. Florea, J.~Anvik, and R.~Andonie, ``Spark-based cluster implementation of
  a bug report assignment recommender system,'' in \emph{International
  Conference on artificial intelligence and soft computing}.\hskip 1em plus
  0.5em minus 0.4em\relax Springer, 2017, pp. 31--42.

\bibitem{fu2017easy}
W.~Fu and T.~Menzies, ``Easy over hard: A case study on deep learning,'' in
  \emph{Proceedings of the 2017 11th joint meeting on foundations of software
  engineering}, 2017, pp. 49--60.

\bibitem{Mikolov2013}
\BIBentryALTinterwordspacing
T.~Mikolov, K.~Chen, G.~Corrado, and J.~Dean, ``Efficient estimation of word
  representations in vector space,'' in \emph{1st International Conference on
  Learning Representations, {ICLR} 2013, Scottsdale, Arizona, USA, May 2-4,
  2013, Workshop Track Proceedings}, Y.~Bengio and Y.~LeCun, Eds., 2013.
  [Online]. Available: \url{http://arxiv.org/abs/1301.3781}
\BIBentrySTDinterwordspacing

\bibitem{guo2020developer}
S.~Guo, X.~Zhang, X.~Yang, R.~Chen, C.~Guo, H.~Li, and T.~Li, ``Developer
  activity motivated bug triaging: via convolutional neural network,''
  \emph{Neural Processing Letters}, vol.~51, no.~3, pp. 2589--2606, 2020.

\bibitem{PenningtonSM14}
\BIBentryALTinterwordspacing
J.~Pennington, R.~Socher, and C.~D. Manning, ``Glove: Global vectors for word
  representation,'' in \emph{Proceedings of the 2014 Conference on Empirical
  Methods in Natural Language Processing, {EMNLP} 2014, October 25-29, 2014,
  Doha, Qatar, {A} meeting of SIGDAT, a Special Interest Group of the {ACL}},
  A.~Moschitti, B.~Pang, and W.~Daelemans, Eds.\hskip 1em plus 0.5em minus
  0.4em\relax {ACL}, 2014, pp. 1532--1543. [Online]. Available:
  \url{https://doi.org/10.3115/v1/d14-1162}
\BIBentrySTDinterwordspacing

\bibitem{lee2022light}
J.~Lee, K.~Han, and H.~Yu, ``A light bug triage framework for applying large
  pre-trained language model,'' in \emph{37th IEEE/ACM International Conference
  on Automated Software Engineering}, 2022, pp. 1--11.

\bibitem{bucilu2006model}
C.~Buciluǎ, R.~Caruana, and A.~Niculescu-Mizil, ``Model compression,'' in
  \emph{Proceedings of the 12th ACM SIGKDD international conference on
  Knowledge discovery and data mining}, 2006, pp. 535--541.

\bibitem{mccloskey1989catastrophic}
M.~McCloskey and N.~J. Cohen, ``Catastrophic interference in connectionist
  networks: The sequential learning problem,'' in \emph{Psychology of learning
  and motivation}.\hskip 1em plus 0.5em minus 0.4em\relax Elsevier, 1989,
  vol.~24, pp. 109--165.

\bibitem{he2020deberta}
P.~He, X.~Liu, J.~Gao, and W.~Chen, ``Deberta: Decoding-enhanced bert with
  disentangled attention,'' \emph{arXiv preprint arXiv:2006.03654}, 2020.

\bibitem{liu2019roberta}
Y.~Liu, M.~Ott, N.~Goyal, J.~Du, M.~Joshi, D.~Chen, O.~Levy, M.~Lewis,
  L.~Zettlemoyer, and V.~Stoyanov, ``Roberta: A robustly optimized bert
  pretraining approach,'' \emph{arXiv preprint arXiv:1907.11692}, 2019.

\bibitem{wu2018empirical}
H.~Wu, H.~Liu, and Y.~Ma, ``Empirical study on developer factors affecting
  tossing path length of bug reports,'' \emph{IET Software}, vol.~12, no.~3,
  pp. 258--270, 2018.

\bibitem{zhang2020efficient}
W.~Zhang, ``Efficient bug triage for industrial environments,'' in \emph{2020
  IEEE International Conference on Software Maintenance and Evolution
  (ICSME)}.\hskip 1em plus 0.5em minus 0.4em\relax IEEE, 2020, pp. 727--735.

\bibitem{xuan2012developer}
J.~Xuan, H.~Jiang, Z.~Ren, and W.~Zou, ``Developer prioritization in bug
  repositories,'' in \emph{2012 34th International Conference on Software
  Engineering (ICSE)}.\hskip 1em plus 0.5em minus 0.4em\relax IEEE, 2012, pp.
  25--35.

\bibitem{bhattacharya2010fine}
P.~Bhattacharya and I.~Neamtiu, ``Fine-grained incremental learning and
  multi-feature tossing graphs to improve bug triaging,'' in \emph{2010 IEEE
  International Conference on Software Maintenance}.\hskip 1em plus 0.5em minus
  0.4em\relax IEEE, 2010, pp. 1--10.

\bibitem{sanh2019distilbert}
V.~Sanh, L.~Debut, J.~Chaumond, and T.~Wolf, ``Distilbert, a distilled version
  of bert: smaller, faster, cheaper and lighter,'' \emph{arXiv preprint
  arXiv:1910.01108}, 2019.

\bibitem{lan2019albert}
Z.~Lan, M.~Chen, S.~Goodman, K.~Gimpel, P.~Sharma, and R.~Soricut, ``Albert: A
  lite bert for self-supervised learning of language representations,''
  \emph{arXiv preprint arXiv:1909.11942}, 2019.

\bibitem{feng2020codebert}
Z.~Feng, D.~Guo, D.~Tang, N.~Duan, X.~Feng, M.~Gong, L.~Shou, B.~Qin, T.~Liu,
  D.~Jiang, and M.~Zhou, ``Codebert: A pre-trained model for programming and
  natural languages,'' 2020.

\bibitem{pythorch}
pytorch.org, ``{TyTorch},'' \url{https://pytorch.org/}, [Online], [Accessed
  10-13-2022].

\bibitem{transformer}
huggingface.co, ``{Transformers},'' \url{https://huggingface.co/}, [Online],
  [Accessed 10-13-2022].

\bibitem{sklearn}
scikit learn.org, ``{scikit-learn | Machine Learning in Python},''
  \url{https://scikit-learn.org/stable/index.html}, [Online], [Accessed
  10-13-2022].

\bibitem{lilja2005measuring}
D.~J. Lilja, \emph{Measuring computer performance: a practitioner's
  guide}.\hskip 1em plus 0.5em minus 0.4em\relax Cambridge university press,
  2005.

\bibitem{mendenhall2013introduction}
W.~Mendenhall, R.~Beaver, and B.~Beaver, \emph{Introduction to Probability and
  Statistics}.\hskip 1em plus 0.5em minus 0.4em\relax Cengage Learning, 2013.

\bibitem{vaswani2017attention}
A.~Vaswani, N.~Shazeer, N.~Parmar, J.~Uszkoreit, L.~Jones, A.~N. Gomez,
  {\L}.~Kaiser, and I.~Polosukhin, ``Attention is all you need,''
  \emph{Advances in neural information processing systems}, vol.~30, 2017.

\end{thebibliography}
\end{document}